%% file: main.tex
\documentclass[12pt, letterpaper, preprint, comicneue]{aastex63}
\usepackage[T1]{fontenc}
\usepackage{fontawesome}
\usepackage{color}
\usepackage{amsmath}
\usepackage{natbib}
\usepackage{ctable}
\usepackage{bm}
\usepackage[normalem]{ulem} 
\usepackage{xspace}
\usepackage{paralist}
\usepackage{fontawesome}

\let\oldbibliography\thebibliography % killin' me.
\renewcommand{\thebibliography}[1]{%
  \oldbibliography{#1}%
  \setlength{\itemsep}{0pt}%
  \setlength{\parsep}{0pt}%
  \setlength{\parskip}{0pt}%
  \setlength{\bibsep}{0ex}
  \raggedright
}
\newcommand{\given}{\,|\,}

\newcommand{\bfi}[1]{\textbf{\textit{#1}}}

\let\oldAA\AA
\renewcommand{\AA}{\text{\normalfont\oldAA}}
\newcommand{\btheta}{\boldsymbol{\theta}}
\newcommand{\bphi}{\boldsymbol{\phi}}

\newcommand{\simbig}{{\sc SimBIG}}
\newcommand{\foreign}[1]{\textsl{#1}}
\newcommand{\etal}{\foreign{et~al.}}

\newcommand{\bitem}{\begin{itemize}}
\newcommand{\eitem}{\end{itemize}}
\newcommand{\beq}{\begin{equation}}
\newcommand{\eeq}{\end{equation}}

\definecolor{orange}{rgb}{1,0.5,0}

\begin{document} \sloppy\sloppypar\frenchspacing 

\title{\simbig: Mock Challenge for a Forward Modeling Approach to Galaxy Clustering}

\newcounter{affilcounter}
\author[0000-0003-1197-0902]{ChangHoon Hahn}
\altaffiliation{changhoon.hahn@princeton.edu.com}
\affil{Department of Astrophysical Sciences, Princeton University, Princeton NJ 08544, USA} 

\author{Michael Eickenberg}
\affil{Center for Computational Mathematics, Flatiron Institute, 162 5th Avenue, New York, NY 10010, USA}

\author{Shirley Ho}
\affil{Center for Computational Astrophysics, Flatiron Institute, 162 5th Avenue, New York, NY 10010, USA}

\author{Jiamin Hou}
\affil{Department of Astronomy, University of Florida, 211 Bryant Space Science Center, Gainesville, FL 32611, USA}
\affil{Max-Planck-Institut f\"ur Extraterrestrische Physik, Postfach 1312, Giessenbachstrasse 1, 85748 Garching bei M\"unchen, Germany}

\author{Pablo Lemos}
\affil{Department of Physics, Universit\'{e} de Montr\'{e}al, Montr\'{e}al, 1375 Avenue Th\'{e}r\`{e}se-Lavoie-Roux, QC H2V 0B3, Canada}
\affil{Mila - Quebec Artificial Intelligence Institute, Montr\'{e}al, 6666 Rue Saint-Urbain, QC H2S 3H1, Canada}
\affil{Center for Computational Mathematics, Flatiron Institute, 162 5th Avenue, New York, NY 10010, USA}

\author[0000-0002-0637-8042]{Elena Massara}
\affil{Waterloo Centre for Astrophysics, University of Waterloo, 200 University Ave W, Waterloo, ON N2L 3G1, Canada}
\affil{Department of Physics and Astronomy, University of Waterloo, 200 University Ave W, Waterloo, ON N2L 3G1, Canada}

\author{Chirag Modi}
\affil{Center for Computational Mathematics, Flatiron Institute, 162 5th Avenue, New York, NY 10010, USA}
\affil{Center for Computational Astrophysics, Flatiron Institute, 162 5th Avenue, New York, NY 10010, USA}

\author[0000-0001-8841-9989]{Azadeh Moradinezhad Dizgah}
\affil{D\'epartement de Physique Th\'eorique, Universit\'e de Gen\`eve, 24 quai Ernest Ansermet, 1211 Gen\`eve 4, Switzerland}

\author[0000-0003-0055-0953]{Bruno R\'egaldo-Saint Blancard}
\affil{Center for Computational Mathematics, Flatiron Institute, 162 5th Avenue, New York, NY 10010, USA}

\author{Muntazir M. Abidi}
\affil{D\'epartement de Physique Th\'eorique, Universit\'e de Gen\`eve, 24 quai Ernest Ansermet, 1211 Gen\`eve 4, Switzerland}

\begin{abstract}
    Simulation-Based Inference of Galaxies (\simbig) is a forward modeling framework 
    for analyzing galaxy clustering using simulation-based inference. 
    In this work, we present the \simbig~forward model, which is designed to match the 
    observed SDSS-III BOSS CMASS galaxy sample.
    The forward model is based on high-resolution {\sc Quijote} $N$-body simulations 
    and a flexible halo occupation model.
    It includes full survey realism and models observational systematics such as 
    angular masking and fiber collisions. 
    We present the ``mock challenge'' for validating the accuracy of 
    posteriors inferred from \simbig~using a suite of 1,500 test
    simulations constructed using forward models with a different $N$-body simulation, 
    halo finder, and halo occupation prescription. 
    As a demonstration of \simbig, we analyze the power spectrum multipoles out
    to $k_{\rm max} = 0.5\,h/{\rm Mpc}$ and infer the posterior of $\Lambda$CDM 
    cosmological and halo occupation parameters.
    %and conduct the mock challenge. 
    Based on the mock challenge, we find that our constraints on $\Omega_m$ 
    and $\sigma_8$ are unbiased, but conservative.
    %Additional simulations can improve these constraints. 
    Hence, the mock challenge demonstrates that \simbig~provides a robust framework 
    for inferring cosmological parameters from galaxy clustering on non-linear scales 
    and a complete framework for handling observational systematics. 
    In subsequent work, we will use \simbig~to analyze summary statistics beyond
    the power spectrum including the bispectrum, marked power spectrum, skew spectrum, 
    wavelet statistics, and field-level statistics. 
\end{abstract} 

\input{intro}

\input{obs}
\input{fm}
\input{test}
\input{results}
\input{discuss}
\input{summary}

\section*{Acknowledgements}
It's a pleasure to thank 
Mikhail M. Ivanov and Yosuke Kobayashi for providing us with the posteriors used for comparison. We also thank Peter Melchior, Uro{\u s}~Seljak, David Spergel, Licia Verde, and Benjamin D. Wandelt for valuable discussions. 
This work was supported by the AI Accelerator program of the Schmidt Futures Foundation. JH has received funding from the European Union’s Horizon 2020 research and innovation program under the Marie Sk\l{}odowska-Curie grant agreement No 101025187. AMD acknowledges funding from Tomalla Foundation for Research in Gravity and Boninchi Foundation. 

%\appendix
%\bibliographystyle{mnras}
\bibliography{simbig} 
\end{document}

%% file: intro.tex
\section{Introduction} \label{sec:intro}
The spatial distribution of galaxies contains key cosmological information about
our Universe. 
From the statistical clustering of galaxies, we can measure the expansion 
history of the Universe and the growth of 
structure~\citep{sargent1977, kaiser1987, eisenstein1998a, hamilton1998, seo2003}.
With these measurements we can probe the nature of dark energy and test
theories of gravity~\citep[\emph{e.g.}][]{jain2013, kim2015, huterer2015}.
Precise measurements of galaxy clustering can also be used to test 
inflation~\citep{liddle2000, dalal2008, slosar2008, ho2015}
and measure the sum of neutrino masses~\citep[][see~\citealt{lesgourgues2013}
for a review]{font-ribera2014, beutler2014a}.  

Upcoming galaxy surveys will collect an unprecedented amount of data to take
advantage of this cosmological information and produce the most precise
measurements of galaxy clustering. 
In particular, spectroscopic galaxy surveys will be conducted using the 
Dark Energy Spectroscopic Instrument~\citep[DESI;][]{desicollaboration2016,
desicollaboration2016a, abareshi2022}, 
the Subaru Prime Focus Spectrograph~\citep[PFS;][]{takada2014, tamura2016}, 
the ESA {\em Euclid} satellite mission~\citep{laureijs2011}, 
and the Nancy Grace Roman Space Telescope~\citep[Roman;][]{spergel2015, wang2022a} 
over the next decade. 
They will precisely measure the three-dimensional clustering of galaxies over
unprecedented cosmic volumes. 
Combined with constraints from other cosmological probes, such as the cosmic
microwave background (CMB), these surveys will produce the most precise 
constraints on the cosmological parameters. 
Moreover, they will precisely test the  standard $\Lambda$CDM model and enable 
us to search for new physics beyond the standard paradigm. 
%these surveys will produce the most precise constraints on the cosmological parameters and stringent tests of the standard $\Lambda$CDM paradigm and enable us to search for new physics beyond the standard paradigm.

The current standard analyses for galaxy clustering use theoretical models of
galaxy clustering from perturbation theory of large-scale 
structure~\citep[PT;][see \citealt{bernardeau2002} and \citealt{desjacques2016}
for a review]{beutler2017, ivanov2020}.
These standard analyses have major limitations. 
First, PT models cannot accurately model non-linear galaxy clustering. 
The primary analyses of the Sloan Digital Sky Survey (SDSS)-III Baryon
Oscillation Spectroscopic Survey~\citep[BOSS;][]{eisenstein2011, dawson2013}
only model the galaxy power spectrum, the two-point clustering statistic, to 
$k_{\rm max} < 0.2\,h/{\rm Mpc}$~\citep[\emph{e.g.}][]{beutler2014a, beutler2017, grieb2017}.
Even recent models that use an effective field theory approach with non-linear biasing, 
IR resummation, and ``counterterms''~\citep[\emph{e.g.}][]{carrasco2012, senatore2014,
senatore2015, perko2016, ivanov2020, damico2022} are limited to 
$k_{\rm max} \sim 0.2\,h/{\rm Mpc}$. 
%For higher-order clustering statistics, PT models struggle on even larger scales. 
For the bispectrum, the lowest higher-order three-point statistic, analyses can 
only include measurements out to $k \le 0.08\,h/{\rm Mpc}$~\citep{philcox2021}.
\cite{damico2022} extend their analysis to $k \sim 0.2\,h/{\rm Mpc}$ for the
bispectrum monopole but they required 33 additional parameters. 

Another limitation is that standard galaxy clustering analyses assume that the
likelihood used in their Bayesian parameter inference has a Gaussian form. 
This assumption relies on the central limit theorem, which breaks down on large
scales with low signal-to-noise and on small scales where modes are highly
correlated. 
\cite{hahn2019c} found that the non-Gaussianity of the
likelihood can significantly bias the overall parameter constraints. 
The effect may be more pronounced at the precision level of upcoming surveys
and for observables beyond the power spectrum.
Lastly, standard galaxy clustering analyses rely on corrections to the
clustering measurements, typically through weights imposed on galaxies, to
account for observational systematics. 
While some systematics can be successfully corrected with weights, others like
fiber collision pose serious challenges for upcoming surveys. 
Improved correction schemes for fiber collisions may be sufficient for power
spectrum analyses~\citep{guo2012, hahn2017a, pinol2017, bianchi2018, smith2019}.
Nevertheless, no correction schemes have yet been designed or demonstrated for higher-order 
statistics. 

Meanwhile, there is a growing consensus among forecasts that there is
significant cosmological information in higher-order statistics and on non-linear scales. 
For example, \cite{hahn2021a} recently used the {\sc Quijote} suite of
simulations to forecast the cosmological information content of the
redshift-space galaxy bispectrum. 
They found that constraints on cosmological parameters $\Omega_m$, $\Omega_b$,
$h$, $n_s$, and $\sigma_8$ improve by more than a factor of 2 when the
bispectrum is included. 
Including smaller scales ($0.2 < k < 0.5\,h/{\rm Mpc}$) also tightens 
constraints by an additional factor of 2. 
Similar forecasts made using other higher-order statistics, \emph{e.g.} the
marked power spectrum~\citep{massara2020, massara2022}, the reconstructed 
power spectrum~\citep{wang2022}, the skew spectra~\citep{hou2022}, wavelet 
statistics~\citep{eickenberg2022}, find consistent improvements over the 
power spectrum. 
Despite the advantages that these observables promise, the limitations above
prevent most of them from being analyzed within the standard galaxy clustering 
analysis framework. 

In this work, we present the Simulation-Based Inference of Galaxies (\simbig),
an alternative approach to analyzing galaxy clustering that addresses each of the
major limitations of standard galaxy clustering analyses. 
First, \simbig~uses simulated forward models of the galaxy distribution. 
As a result, it can accurately model galaxy clustering on non-linear scales 
beyond the limits of PT.  
Furthermore, we can analyze any observable (\emph{i.e.} summary statistic) 
that can be measured in the observed galaxy distribution. 
Second, \simbig~uses simulation-based inference (SBI; also known as
``likelihood-free inference'' or ``implicit likelihood inference'') 
that enables rigorous inference using only a forward model of the observables.  
It makes no assumptions on the functional form of the likelihood.
It leverages the machine learning based density estimation to accurately describe
even high-dimensional posteriors efficiently with a limited number of
simulations~(see \citealt{cranmer2020} and references therein). 
SBI has already been used for Bayesian inference in large-scale structure
studies~\citep[\emph{e.g.}][]{hahn2017b, alsing2019, hassan2021, jeffrey2021,
makinen2021, lemos2022}. 
It has also been widely adopted in broader astronomical
contexts~\citep{dax2021, huppenkothen2021, lemos2021, tortorelli2021, zhang2021, 
hahn2022a, lemos2022}.
Lastly, \simbig~provides a more complete framework for accounting for 
observational systematics. 
Unlike the conventional approach of using weights to correct for systematics, 
the forward model in \simbig~includes the systematics. 
Some observational systematics are often much easier to model than to correct 
in the clustering measurements and by forward modeling them a correction scheme
does not need to be developed for every summary statistic. 
Altogether, \simbig~provides a framework that can robustly analyze galaxy
clustering down to non-linear scales and with higher-order statistics. 

In this work and in an accompanying paper~\citep[][hereafter H22a]{simbig_letter}, we demonstrate
\simbig~by applying it to BOSS.
As the first papers of a series, we apply \simbig~to the power spectrum multipoles,
$P_\ell(k)$, so that we can validate the \simbig~framework in detail and make
comparisons to previous galaxy clustering studies in the literature. 
H22a presents the cosmological constraints and briefly
describes \simbig. 
In this work, we describe the details of \simbig~pipeline: the forward model
and the SBI framework. 
We also present the mock challenge for validating the accuracy of the 
\simbig~posteriors using a suite of test simulations constructed with 
different forward models.
In subsequent works, we will apply \simbig~to summary statistics beyond the
power spectrum including the bispectrum, the marked power spectrum, the 
skew spectra, the void probability function, and wavelet statistics.
We will also analyze field-level summary statistics that compress all the 
information in the galaxy field using convolutional and graph neural networks.

We begin in Section~\ref{sec:obs} by describing the observational galaxy sample
that we analyze with \simbig.
Then, we present the details of the \simbig~pipeline in
Section~\ref{sec:simbig}.
Afterwards, in Section~\ref{sec:tests}, we describe how we construct the suite
of test simulations used for the mock challenge. 
We present the results of the mock challenge in Section~\ref{sec:results} and
discuss our findings in Section~\ref{sec:discuss}.

%% file: obs.tex
\begin{figure}
\begin{center}
    \includegraphics[width=0.9\textwidth]{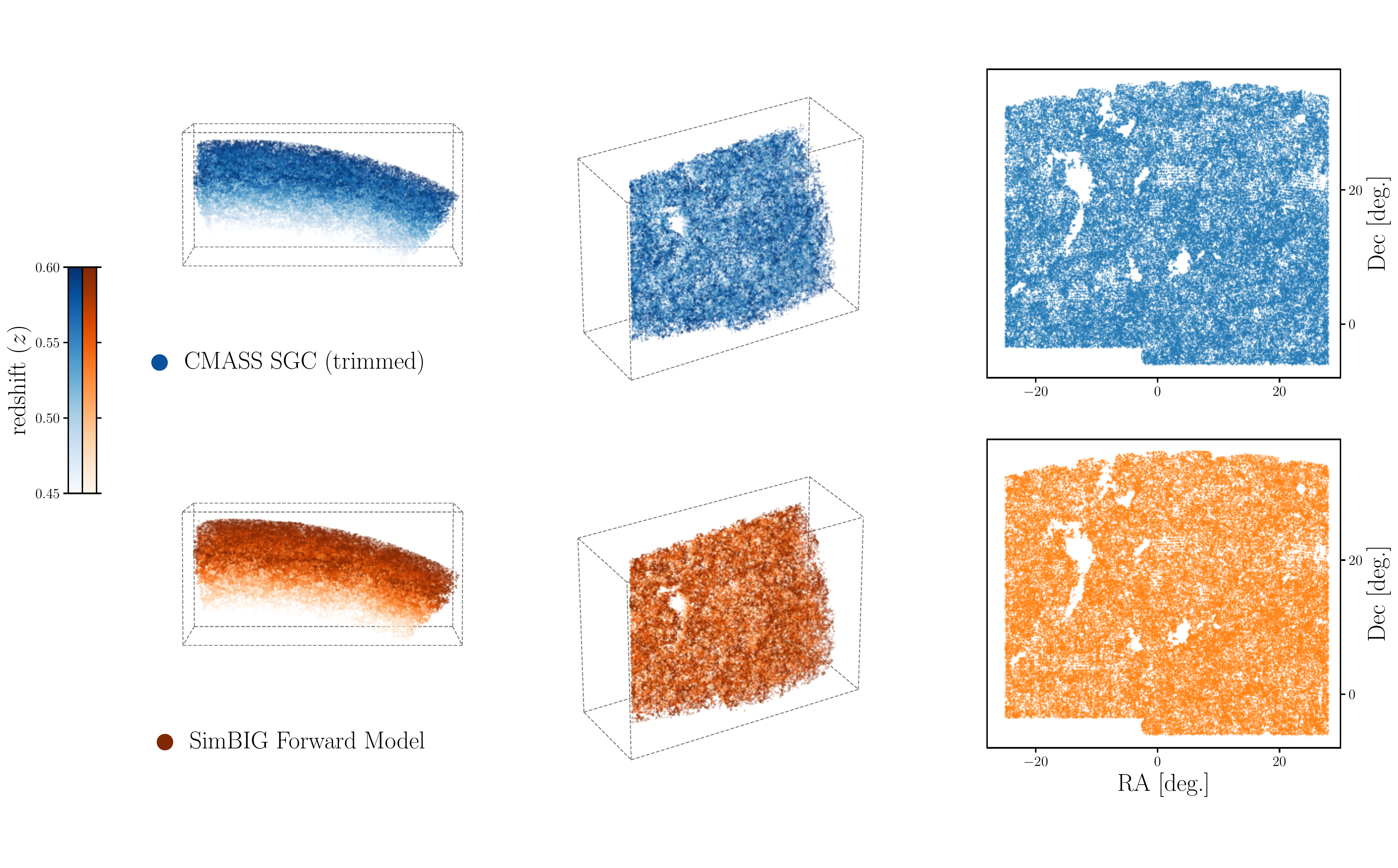}
    \caption{
        Comparison of the observed CMASS-SGC galaxy sample (top; blue) with
        the \simbig~forward model (bottom; orange).  
        We present the 3D distribution of galaxies at two different viewing 
        angles in the left and center panels.
        The right panels plot the angular distribution of galaxies. 
        Our CMASS-SGC galaxy sample contains 109,636 galaxies. 
        Our forward model constructs realistic CMASS-like galaxy samples from
        {\sc Quijote} $N$-body simulations using an HOD model that includes the
        effects of assembly bias, concentration bias, and velocity biases for
        central and satellite galaxies. 
        It also applies observational systematics: the CMASS survey geometry,
        veto masks, and fiber collisions.
        Hence, our forward modeled galaxy catalogs have a selection function
        that closely matches the selection function of CMASS.  
        For more comparisons of the 3D distributions, we refer readers to
        \href{https://youtube.com/playlist?list=PLQk8Faa2x0twK3fgs55ednnHD2vbIzo4z}{\faGlobe}.% \footnote{\url{https://youtube.com/playlist?list=PLQk8Faa2x0twK3fgs55ednnHD2vbIzo4z}}.
        \label{fig:fm_demo}
    }
\end{center}
\end{figure}

\section{Observations: BOSS CMASS Galaxies} \label{sec:obs}
We apply \simbig~to the publicly available galaxy
catalog\footnote{\url{https://data.sdss.org/sas/dr12/ boss/lss/}}
from the Sloan Digital Sky Survey (SDSS)-III Baryon Oscillation Spectroscopic
Survey (BOSS) Data Release 12~\citep{eisenstein2011, dawson2013}.
BOSS observed two galaxy samples, LOWZ and CMASS, that span the redshift ranges
$0.2 \lesssim z \lesssim 0.5$ and $0.43 < z < 0.7$.
In this work, we focus on the CMASS galaxy sample, which consists of
Luminous Red Galaxies (LRGs) with a prominent $4000\AA$ break in their
spectral energy distribution for reliable redshift measurements.
CMASS LRGs reside in massive halos with $M_h \gtrsim 10^{13}M_\odot$ and the
sample has a comoving galaxy number density of 
$\bar{n} \approx 3\times 10^{-4}\, h^3{\rm Mpc}^{-3}$~\citep{white2011,
leauthaud2016, saito2016, reid2016, zhai2017}.

Each CMASS galaxy is assigned a set of weights to account for incompleteness 
and observational systematics in BOSS.
These systematics include redshift failures, fiber collisions, and variations in
the target density caused by stellar density and seeing conditions. 
We refer readers to \cite{ross2012, ross2017} for more details on the
observational systematics in BOSS. 

In this work, we further restrict our analysis to CMASS galaxies in the
Southern Galactic Cap (SGC) that have ${\rm Dec} > -6$ deg. and 
$-25 < {\rm RA} < 28$ deg. 
We also impose a redshift cut of $0.45 < z < 0.6$. 
As we discuss later in Section~\ref{sec:fm}, we impose these extra
restrictions on the CMASS sample due to the limited volume of the simulations
used in our forward model. 
In the top panels of Figure~\ref{fig:fm_demo}, we present the spatial and
angular distributions of the CMASS-SGC sample (blue).
The left and center panels present the 3D distribution of CMASS-SGC
galaxies at two different viewing angles.
The right panel presents the angular distribution. 
Figure~\ref{fig:fm_demo} illustrates the extra selections that we impose and 
the overall distribution of the 109,636 CMASS-SGC galaxies. 

%% file: fm.tex
\newpage 
\section{Simulation-Based Inference of Galaxies (\simbig)} \label{sec:simbig}

%%%%%%%%%%%%%%%%%%%%%%%%%%%%%%%%%%%%%%%%%%%%%%%%%%%%%%%%%%%%%%%%%%%%%%%%%%%%
% forward model  
%%%%%%%%%%%%%%%%%%%%%%%%%%%%%%%%%%%%%%%%%%%%%%%%%%%%%%%%%%%%%%%%%%%%%%%%%%%%
\subsection{Forward Model} \label{sec:fm}
For our forward model, we start with the $N$-body simulations from
the {\sc Quijote} suite~\citep{villaescusa-navarro2020}.  
In particular, we use the set of high-resolution $\Lambda$CDM Latin-hypercube
(LHC) simulations. 
The simulations are run using the TreePM {\sc Gadget-III} code.
They are constructed using $1024^3$ cold dark matter (CDM) particles
initialized at $z=127$ using 2LPT and gravitationally evolved until $z=0.5$.  
The simulations have a cosmological volume of $1 (h^{-1}{\rm Gpc})^3$.  
By using $N$-body simulations, instead of more approximate schemes such as
particle mesh, we can accurately model the clustering of matter down to small,
non-linear scales. 
%\cite{villaescusa-navarro2020} verified that the clustering in the $N$-body  simulations are accurate to $k\gtrsim0.5\,h/{\rm Mpc}$.

In this work, our goal is to forward model the spatial distribution of galaxies. 
%With a PT approach, this is typically done using a galaxy bias model. 
We use the halo occupation distribution~\citep[HOD;][]{berlind2002, zheng2007} 
framework to construct galaxy samples from the $N$-body simulations. 
First, we identify dark matter halos from the simulations using the 
{\sc Rockstar} halo finder~\citep{behroozi2013a}.
{\sc Rockstar} identifies halos using phase space information of the dark matter
particles in the simulation and has been shown to accurately determine the 
location of halos and resolve their substructure~\citep{knebe2011}. 

%\begin{figure}
%\begin{center}
%    \includegraphics[width=0.5\textwidth]{figs/hod_demo.pdf}
%    \caption{\label{fig:hod}
%    For the \simbig~forward model, we use an HOD model that extends the standard 
%    Z07 model with assembly, concentration, and velocity biases. 
%    To illustrate our HOD model, we present the halo occupation as a function 
%    of $M_h$ for halos with low (blue) and high (orange) concentration for 
%    $A_{\rm bias} = 0.5$ and the Z07 parameters set to the best-fit values 
%    from \cite{reid2014} (black dotted). 
%    We also mark the full range of halo occupation produced by varying our 
%    HOD parameters within the prior range in gray shaded region.
%    }
%\end{center}
%\end{figure}

Next, we populate the dark matter halos with galaxies using a
state-of-the-art HOD model that includes assembly, concentration, and velocity
biases. 
An HOD model provides the statistical prescription for determining the number
of galaxies as well as their positions and velocities within the halo based on 
its properties.  
Our HOD model is based on the standard \cite{zheng2007} model (hereafter Z07). 
In this model, the number of central and satellite galaxies in a halo is
determined by the mass of the halo, $M_h$, and five free HOD parameters: 
$\log M_{\rm min}, \sigma_{\log M}, \log M_0, \log M_1, \alpha$ (see 
Table~\ref{tab:prior} for a brief description of each parameter). 
Central galaxies are placed at the center of the halos and with the same
velocity as the halo.  
The positions and velocities of satellite galaxies are sampled according to a
NFW~\citep{navarro1997} profile. 
%Mention halo mass resolution of high-res Quijote sims (~3x10^12) and that it matches well CMASS galaxies

%In the Z07 HOD, galaxy occupation is solely based on $M_h$.  
While the Z07 HOD model has been shown to successfully reproduce the 
clustering of CMASS galaxies~\citep{white2011, reid2014, manera2015},
a growing number of works now suggest that galaxies occupy halos 
in ways that depend on halo properties beyond $M_h$, such as the assembly 
history of halos~\citep{gao2005, wechsler2006, zentner2007, 
dalal2008, lacerna2011, miyatake2016, more2016, zentner2016, 
vakili2019, hadzhiyska2021, hadzhiyska2022a}.
We include this ``assembly bias'' effect by supplementing the Z07 model using a
decorated HOD prescription~\citep{hearin2016}. 
For a given $M_h$, halos are split into two bins based on their concentrations. 
The high- and low-concentration halos are assigned different numbers of central 
and satellite galaxies 
but the average occupation of both populations is the same as the Z07 model. 
The difference is controlled by $A_{\rm bias}$, a parameter that ranges from -1 
to 1.
For more positive values of $A_{\rm bias}$, high-concentration halos host more 
galaxies relative to low-concentration ones of the same mass.
With this prescription, our HOD depends not only on $M_h$ but also on the halo
concentration, which is a proxy for its assembly history~\citep{mao2015}. 
%We demonstrate the decorated HOD in Figure~\ref{fig:hod}, where we
%present the halo occupation number (average number of galaxies in the halo)
%as a function of $M_h$ for halos with low (blue) and high (orange) 
%concentration for $A_{\rm bias} = 0.5$.
%The five Z07 HOD parameters are set to the best-fit values from \cite{reid2014}
%derive from  fitting the two-point correlation function of CMASS galaxies 
%(black dotted).

We add extra flexibility to our HOD model by including concentration and
velocity biases.
Concentration bias allows the concentration of satellites galaxies to deviate
from the NFW profile of their halos. 
We parameterize concentration bias using the free parameter $\eta_{\rm conc}$,
which sets the ratio between the concentration of satellites and halos
$c_{\rm sat}/c_{\rm halo}$. 
Meanwhile, central and satellite velocity biases rescale the velocity of
central and satellite galaxies with respect to the host halo. 
We use parameters $\eta_{\rm cen}, \eta_{\rm sat}$ that set the velocity
dispersions of central and satellite galaxies: 
$\sigma_{\rm cen} = \eta_{\rm cen} \sigma_{\rm halo}$ and 
$\sigma_{\rm sat} = \eta_{\rm sat} \sigma_{\rm halo}$, where 
$\sigma_{\rm halo}$ is the velocity dispersion of the dark matter halos.
In total, we use an HOD model with 9 free parameters, which we list in
Table~\ref{tab:prior}. 
We note that this is similar to the HOD model used in~\cite{zhai2022}.

From the HOD galaxy catalog in a $1\,(h^{-1}{\rm Gpc})^3$ box, we construct a
CMASS-like galaxy catalog by applying the survey geometry and observational
systematics. 
We first remap the simulation box to a cuboid volume with dimensions 
$1.414 \times 1.224 \times 0.5773\,(h^{-1}{\rm Gpc})^3$ using the 
\cite{carlson2010} method, which is one-to-one volume preserving and keeps
local structures intact. 
Moreover, it allows us to efficiently fit the CMASS survey geometry. 
After translating and rotating the cuboid, we cut out the survey geometry from
it using {\sc Mangle} polygons~\citep{swanson2008} from BOSS~\citep{dawson2013}
that includes the angular footprint of the survey as well as the veto mask,
which includes masking for bright stars, centerpost, bad field, collision
pen 
priority. 
We then trim the forward modeled galaxy catalog with the same $0.45 < z < 0.6$
range that we imposed on the observations. 
Lastly, we apply fiber collisions.
We identify all pairs of galaxies within an angular scale of
$62^{\prime\prime}$; then, for a randomly selected 60\% of the pairs, we remove
one of the galaxies from the sample. 

In Figure~\ref{fig:fm_demo}, we present a comparison of the CMASS sample
(top; blue) with a galaxy catalog generated from our forward model (bottom; orange). 
In the left and center panels, we present the 3D distribution of galaxies at 
two different viewing angle.
For more comparisons of the 3D distributions, we refer readers to
\href{https://youtube.com/playlist?list=PLQk8Faa2x0twK3fgs55ednnHD2vbIzo4z}{\faGlobe}
\footnote{\url{https://youtube.com/playlist?list=PLQk8Faa2x0twK3fgs55ednnHD2vbIzo4z}}.
In the right panels, we present the angular distribution of the galaxies.  
The forward modeled galaxy catalog has the same detailed angular footprint and
redshift range as the CMASS sample with similar systematics.  

%%%%%%%%%%%%%%%%%%%%%%%%%%%%%%%%%%%%%%%%%%%%%%%%%%%%%%%%%%%%%%%%%%%%%%%%%%%%
%Priors 
%%%%%%%%%%%%%%%%%%%%%%%%%%%%%%%%%%%%%%%%%%%%%%%%%%%%%%%%%%%%%%%%%%%%%%%%%%%%
\begin{table} 
\caption{Parameters of our forward model and their priors.} 
\begin{center}
    \begin{tabular}{ccc} \toprule
        name & description & prior \\[3pt]
        \hline 
        \multicolumn{3}{c}{Cosmological parameters} \\
        \hline 
        $\Omega_m$  & matter energy density     & $\mathcal{U}(0.1, 0.5)$ \\
        $\Omega_b$  & baryon energy density     & $\mathcal{U}(0.03, 0.07)$ \\
        $h$         & dimensionless Hubble constant & $\mathcal{U}(0.5, 0.9)$ \\
        $n_s$       & spectral index of the primordial power spectrum & 
            $\mathcal{U}(0.8, 1.2)$ \\
        $\sigma_8$  & amplitude of matter fluctuations on $8 h^{-1}{\rm Mpc}$ 
        scales & 
            $\mathcal{U}(0.6, 1.0)$ \\[3pt]
        \hline
        \multicolumn{3}{c}{Halo occupation parameters}\\
        \hline 
        $\log M_{\rm min}$ & characteristic mass scale for halos to host a
        central galaxy & 
            $\mathcal{U}(12., 14.)$ \\
        $\sigma_{\log M}$  & scatter of halo mass at fixed galaxy luminosity &
            $\mathcal{U}(0.1, 0.6)$ \\
        $\log M_0$         & minimum halo mass for halos to host a satellite
        galaxy & 
            $\mathcal{U}(13., 15.)$ \\
        $\log M_1$         & characteristic mass scale for halos to host a
        satellite galaxy & 
            $\mathcal{U}(13., 15.)$ \\
        $\alpha$           & power-law index for the mass dependence of
        satellite occupation & 
            $\mathcal{U}(0.0, 1.5)$ \\
        $A_{\rm bias} $    & assembly bias & 
            $\mathcal{N}(0, 0.2)$ over [-1, 1] \\
        $\eta_{\rm conc} $ & concentration bias of satellites & 
            $\mathcal{U}(0.2, 2.0)$ \\
        $\eta_{\rm cen} $  & velocity bias of centrals & 
            $\mathcal{U}(0.0, 0.7)$ \\
        $\eta_{\rm sat} $  & velocity bias of satellites & 
            $\mathcal{U}(0.2, 2.0)$ \\[3pt]
        \hline
        \multicolumn{3}{c}{Nuisance parameters}\\
        \hline
        $A_{\rm shot} $     & $P_0$ shot noise correction & 
            $\mathcal{U}(-10^4, 10^4)$ \\[3pt]
        \hline            
\end{tabular} \label{tab:prior}
\end{center}
\end{table}

\subsection{Priors} \label{sec:prior} 
For the cosmological parameters, $\{\Omega_m, \Omega_b, h, n_s, \sigma_8\}$, we
use uniform priors over the parameter ranges that fully encompass the 
{\em Planck} priors.
We impose these priors by constructing the training data from {\sc Quijote}
$N$-body simulations generated using cosmological parameters in an LHC
configuration.
For the HOD parameters, we choose conservative priors that can produce a broad
range of galaxy populations. 
In particular, for the standard Z07 HOD parameters, 
$\{\log M_{\rm min}, \sigma_{\log M}, \log M_0,  \log M_1, \alpha\}$, 
we choose their prior ranges so that a galaxy sample with a comparable
number of galaxies as the observations can be forward modeled from any 
of the $N$-body simulations in the LHC by using a set of HOD parameter 
values within the priors. 
For instance, this led us to adopt broad conservative priors for 
$\log M_{\rm min}$ and $\sigma_{\log M}$.
For assembly bias, we use a Gaussian prior centered at 0 with $\sigma = 0.2$
over the range [-1, 1] for $A_{\rm bias}$. 
Lastly, for concentration, central velocity, and satellite velocity bias we use
uniform priors on $\eta_{\rm conc}, \eta_{\rm cen}, and \eta_{\rm sat}$,
respectively, with the same range as in \cite{zhai2022}. 
We list priors for our cosmological and HOD parameters in
Table~\ref{tab:prior}. 
%We also present the full range of halo occupation numbers for our HOD models based on the HOD prior range in Figure~\ref{fig:hod}. 

%%%%%%%%%%%%%%%%%%%%%%%%%%%%%%%%%%%%%%%%%%%%%%%%%%%%%%%%%%%%%%%%%%%%%%%%%%%%
% summary statistic  
%%%%%%%%%%%%%%%%%%%%%%%%%%%%%%%%%%%%%%%%%%%%%%%%%%%%%%%%%%%%%%%%%%%%%%%%%%%%
\subsection{Summary Statistic} \label{sec:fm}
The \simbig~framework enables us to derive robust cosmological constraints 
using any summary statistics of the observed galaxy distribution. 
In this work, however, our primary goal is to demonstrate and validate the
\simbig~framework so we use the galaxy power spectrum multipoles, $P_\ell(k)$,
as our summary statistic. 
$P_\ell(k)$ is a standard cosmological observable that has been extensively
analyzed in the literature. 
Later, we compare the constraints that we infer from our analysis to some of
these previous works~\citep{ivanov2020, kobayashi2021}. 

For both observed and forward modeled galaxy samples, we use the
\cite{hand2017a} $P_\ell$ estimator implemented in the
$\mathtt{nbodykit}$ python
package\footnote{\url{https://nbodykit.readthedocs.io/en/latest/index.html}}~\citep{hand2018}.
The estimator is Fast Fourier Transform (FFT) based and uses FFTs with grid
size $N_{\rm grid} = 360$. 
The estimator accounts for the survey geometry using a random catalog that has the same
radial and angular selection functions as the observed catalog but with a much
larger number of objects (>4,000,000) with random angular and radial positions.

When measuring $P_\ell$, we include FKP weights~\citep{feldman1994} with $P_0 =
10^4$ to reduce the variance of the measured $P_\ell$ for a galaxy sample with
non-uniform completeness. 
For the observed galaxy sample, we also include systematic weights. 
For galaxy $i$, we assign $w_{g, i} = w_{{\rm sys},i} w_{{\rm noz}, i}$, where 
$w_{{\rm sys},i}$ is an angular systematic weight based on stellar density and
seeing conditions and $w_{{\rm noz}, i}$ is a redshift failure weight. 
We exclude weights for fiber collisions, which are typically included in other
BOSS analyses \citep[\emph{e.g.}][]{beutler2017}, because we include the effect
in our forward model.  
%Meanwhile, we include $w_{{\rm sys},i}$ and $w_{{\rm noz}, i}$ since these systematics are not included in our forward model. 
We forward model fiber collisions because the standard weighting scheme does 
not accurately correct for them~\citep{hahn2017a}.  
Meanwhile, the other weights successfully account for their corresponding
systematic effects so they are not included~\citep{ross2012, anderson2014}.

For our summary statistic, we include the power spectrum monopole, quadrupole, and
hexadecapole ($\ell$ = 0, 2, and 4).
Each multipole is measured with $\Delta k = 0.005$ and out to 
$k_{\rm max} = 0.5\,h/{\rm Mpc}$ with 100 $k$-bins. 
In addition to the power spectrum, we also include $\bar{n}_g$, the average 
galaxy number density. 
In total, our summary statistic has 301 elements. 

%%%%%%%%%%%%%%%%%%%%%%%%%%%%%%%%%%%%%%%%%%%%%%%%%%%%%%%%%%%%%%%%%%%%%%%%%%%%
% training data 
%%%%%%%%%%%%%%%%%%%%%%%%%%%%%%%%%%%%%%%%%%%%%%%%%%%%%%%%%%%%%%%%%%%%%%%%%%%%
\subsection{Training Data} \label{sec:training}
The SBI in \simbig~requires a training dataset of $(\btheta', \bfi{x}')$ pairs,
where $\btheta'$ is a set of parameter values drawn from the prior and 
$\bfi{x}'$ is some observable --- in our case $P_\ell$ and $\bar{n}_g$ ---
forward modeled using $\btheta'$.
To construct this training dataset, we begin with the 2000 {\sc Quijote}
$N$-body simulations in the LHC configuration.  
For each simulation, we forward model 10 CMASS-like galaxy catalogs using unique
HOD parameters randomly sampled from the prior. 
Afterwards, we measure $P_\ell(k)$ for all of the galaxy catalogs. 

We supplement our training data with an additional parameter: $A_{\rm shot}$, a
nuisance parameter to marginalize over the residual shot noise contribution
beyond the Poisson shot noise. 
We include $A_{\rm shot}$ mainly to be consistent with prior $P_\ell(k)$
analyses~\citep{beutler2017, ivanov2020, kobayashi2021}.
In practice, we include $A_{\rm shot}$ for each $P_0$ by adding a 
constant sampled from a uniform prior: 
$A'_{\rm shot} \sim \mathcal{U}(-10^4, 10^4)$. 
We use the same range for $A_{\rm shot}$ as \cite{kobayashi2021}. 
%$A_{\rm shot}$ also migitates any potential impact of the halo mass limit on $P_0$; we discuss this further in Section~\ref{sec:discuss}. 

\begin{figure}
\begin{center}
    \includegraphics[width=0.9\textwidth]{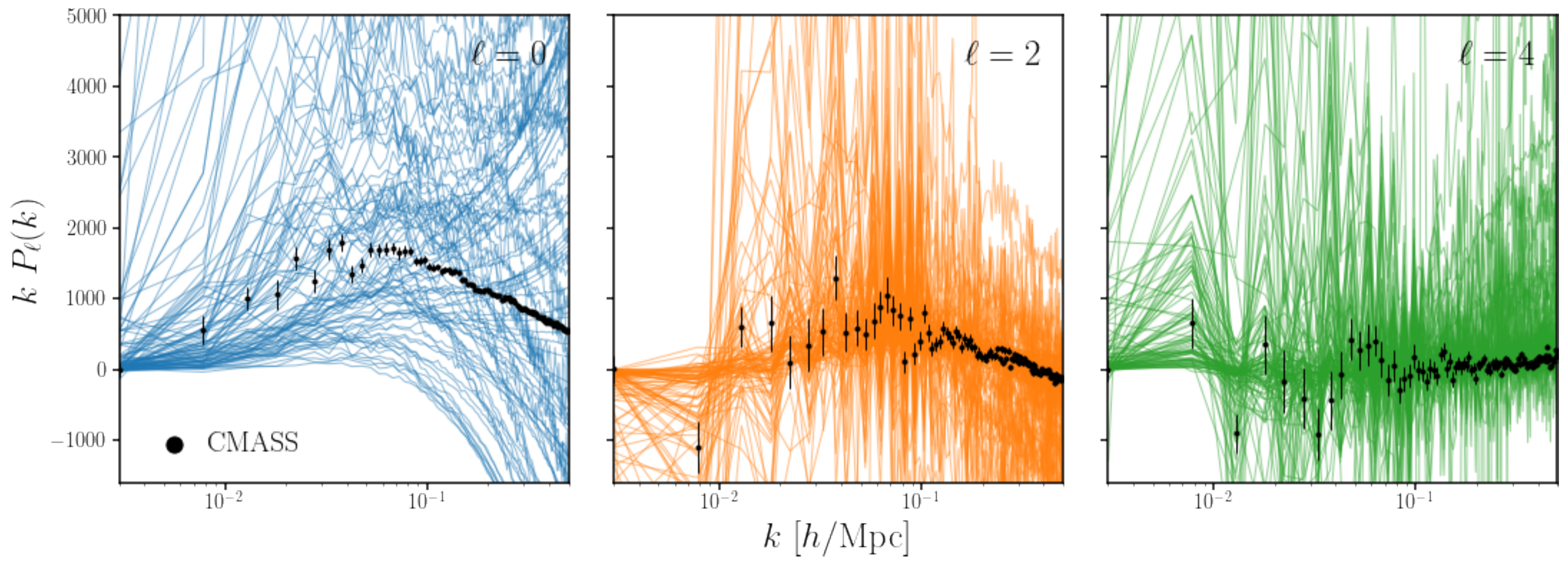}
    \caption{
        Power spectrum multipoles, $k\,P_\ell(k)$, of the simulated galaxy  catalogs
        in our training dataset constructed from the {\sc Quijote} $N$-body 
        simulations using our forward model.
        We randomly select 100 out of the 20,000 catalogs.
        We present the power spectrum monopole, quadrupole, hexadecapole 
        ($\ell = 0, 2, 4$) in the left, center, and right panels. 
        For reference, we include $k\,P_\ell(k)$ measured from the observed 
        CMASS sample (black) with errorbars estimated from the 
        $\mathtt{TEST0}$ simulations.
    }\label{fig:plk}
\end{center}
\end{figure}

In total, we construct a training dataset of 20,000 $(\btheta', P'_\ell)$ pairs.
We present $k\,P_\ell(k)$ for a randomly selected subset of the training dataset in
Figure~\ref{fig:plk}. 
The left, center, and right panels present the monopole, quadrupole, and
hexadecapole respectively.
For reference, we include $P_\ell$ of the observed CMASS sample (black) with 
uncertainties estimated using the $\mathtt{TEST0}$ simulations, which we describe
later in Section~\ref{sec:quij_test}. 
With our choice of conservative priors (Table~\ref{tab:prior}), $P_\ell$
of the training dataset has a broad range that fully encompasses the observed
$P_\ell$. 

%%%%%%%%%%%%%%%%%%%%%%%%%%%%%%%%%%%%%%%%%%%%%%%%%%%%%%%%%%%%%%%%%%%%%%%%%%%%
% training data 
%%%%%%%%%%%%%%%%%%%%%%%%%%%%%%%%%%%%%%%%%%%%%%%%%%%%%%%%%%%%%%%%%%%%%%%%%%%%
\subsection{Simulation-Based Inference with Normalizing Flows} \label{sec:anpe_train}
Our main goal is to infer posterior distributions of cosmological parameters given
a summary statistic of the observations: $p(\btheta\given\bfi{x})$. 
In the standard approach, the posterior is estimated using Markov Chain Monte Carlo 
(MCMC) sampling methods, where the posterior value of each sample is evaluated 
using Bayes' rule based on the likelihood and prior. 
The likelihood is assumed to have a Gaussian functional form: 
$\ln p(\bfi{x}\given\btheta) = -\frac{1}{2}\left(\bfi{x} - m(\btheta)\right)^T
{\bf C}^{-1} \left(\bfi{x} - m(\btheta)\right)$, 
where $m(\btheta)$ is the theoretical model and ${\bf C}$ is the covariance
matrix of the observables. 
MCMC requires drawing a large number of samples from the posterior distribution, 
which precludes simulated forward models from being used for $m(\btheta)$. 
As we discuss in Section~\ref{sec:intro}, this dramatically restricts the 
observables and the physical scales that we can analyze. 
Furthermore, the Gaussianity of the likelihood breaks down on low
signal-to-noise regimes and higher-order statistics~\citep{scoccimarro2000, hahn2019c}. 

SBI offers an alternative that allows us to exploit forward models and 
relaxes the assumptions on the form of the likelihood~\citep{cranmer2020}. 
In this work, we use an SBI method that uses a training dataset to estimate 
the posterior using density estimation~\citep[\emph{e.g.}][]{alsing2018, wong2020, huppenkothen2021,
zhang2021, hahn2022a}.
We use simulated $(\btheta', \bfi{x}')$ pairs to train a neural density 
estimator with parameters $\bphi$: $q_{\bphi}(\btheta \given \bfi{x}')$.
For our density estimator we use ``normalizing flow'' models~\citep{tabak2010, tabak2013}
that use an invertible bijective transformation, $f$, to map a complex target 
distribution to a simple base distribution, $\pi(\bfi{z})$, that is fast to
evaluate.
In our case, the target distribution is the posterior and we use a multivariate 
Gaussian for our base distribution. 
The transformation $f$ is defined to be invertible and have a tractable Jacobian 
so that the target distribution can be evaluated from $\pi(\bfi{z})$ by change of 
variables. 
Since $\pi(\bfi{z})$ is easy to evaluate, we can also easily evaluate the
target distribution.  
A  (NN) is trained to obtain $f$, which provides an extremely flexible
mapping from the base distribution.
In this work, we use Masked Autoregressive
Flow~\citep[MAF;][]{papamakarios2017} models implemented in the $\mathtt{sbi}$
Python
package\footnote{\url{https://github.com/mackelab/sbi/}}~\citep{greenberg2019,
tejero-cantero2020}.
MAF model stacks multiple Masked Autoencoder for Distribution
Estimation~\citep[MADE;][]{germain2015} models to combine normalizing flows
with an autoregressive design~\citep{uria2016} that  is well-suited for estimating 
conditional probability distributions such as a posterior. 

Our goal is to train a normalizing flow that best approximates the posterior: 
$p(\btheta\given\bfi{x}) \approx q_{\bphi}(\btheta\given\bfi{x})$.
First, we split the training data into a training and validation set with a 90/10 
split. 
Then we minimize the KL divergence between 
$p(\btheta, \bfi{x}) = p(\btheta\given\bfi{x}) p(\bfi{x})$ and
$q_{\bphi}(\btheta\given\bfi{x}) p(\bfi{x})$, by maximizing the total log-likelihood  
$\sum_i \log q_{\bphi}(\btheta_i\given \bfi{x}_i)$ over  training set.
We use the {\sc Adam} optimizer~\citep{kingma2017} with a learning rate of $5\times10^{-4}$. 
We prevent overfitting by evaluating the total log-likelihood on the validation
data at every training epoch and stop the training when the validation log-likelihood 
fails to increase after 20 epochs.  
We determine the architecture of our normalizing flow through experimentation. 
We train multiple flows with randomly selected architectures and examine their 
validation losses as a function of epoch to ensure they were appropriately trained. 
Afterwards, we select one of the normalizing flows based on validation loss. 
We note that flows with comparable validation losses infer overall consistent 
posteriors with $\lesssim 5\%$ variation.
The flow we use has 6 MADE blocks, 9 hidden layers, and 186 units.

In $q_{\bphi}(\btheta\given\bfi{x})$, $\btheta$ represents the 15 cosmological, HOD, and 
nuisance parameters and $\bfi{x}$ represents the 301 element summary statistic, $P_\ell(k)$ 
and $\bar{n}_g$.
In principle, we can compress $\bfi{x}$ to reduce the dimensionality of the posterior. 
A variety of compression schemes can be used in conjunction with SBI. 
\cite{alsing2018} and  \cite{charnock2018}, for example, proposed compression schemes 
that maximize the Fisher information. 
These schemes require derivatives of the summary statistic with respect to the parameters 
to calculate the Fisher matrix.
In our case, we use the {\sc Quijote} simulations in an LHC so we do not have access
to derivatives with respect to cosmological parameters.
An alternative compression scheme would be to train a NN that takes in 
the summary statistics as input and predicts the parameters. 
The predicted parameters would serve as the compressed summary statistic. 
We use this regression network to compress $\bfi{x}$ and conduct SBI on the compressed 
statistics.
For our $P_\ell$ analysis, we find no significant difference between using the full 
versus compressed $\bfi{x}$.
We therefore use the full summary statistic in $q_{\bphi}(\btheta\given\bfi{x})$ to infer 
the posteriors. 

%% file: test.tex
\section{Test Simulation Suites} \label{sec:tests}
In this work we validate the \simbig~framework and demonstrate that we can use
it to infer accurate posteriors. 
One way to validate our posteriors is to compare them to previous constraints
in the literature. 
We do this in H22a and later in Section~\ref{sec:results}. 
However, a more rigorous validation of \simbig~is to demonstrate that we can
infer accurate and unbiased parameter constraints for a suite of realistic test
simulations that make different assumptions than our training set. 
In this section, we describe how we construct three different sets of test
simulations: two using {\sc Quijote} and a third using the {\sc AbacusSummit}
$N$-body simulations~\citep{maksimova2021}. 

\begin{figure}
\begin{center}
\includegraphics[width=0.9\textwidth]{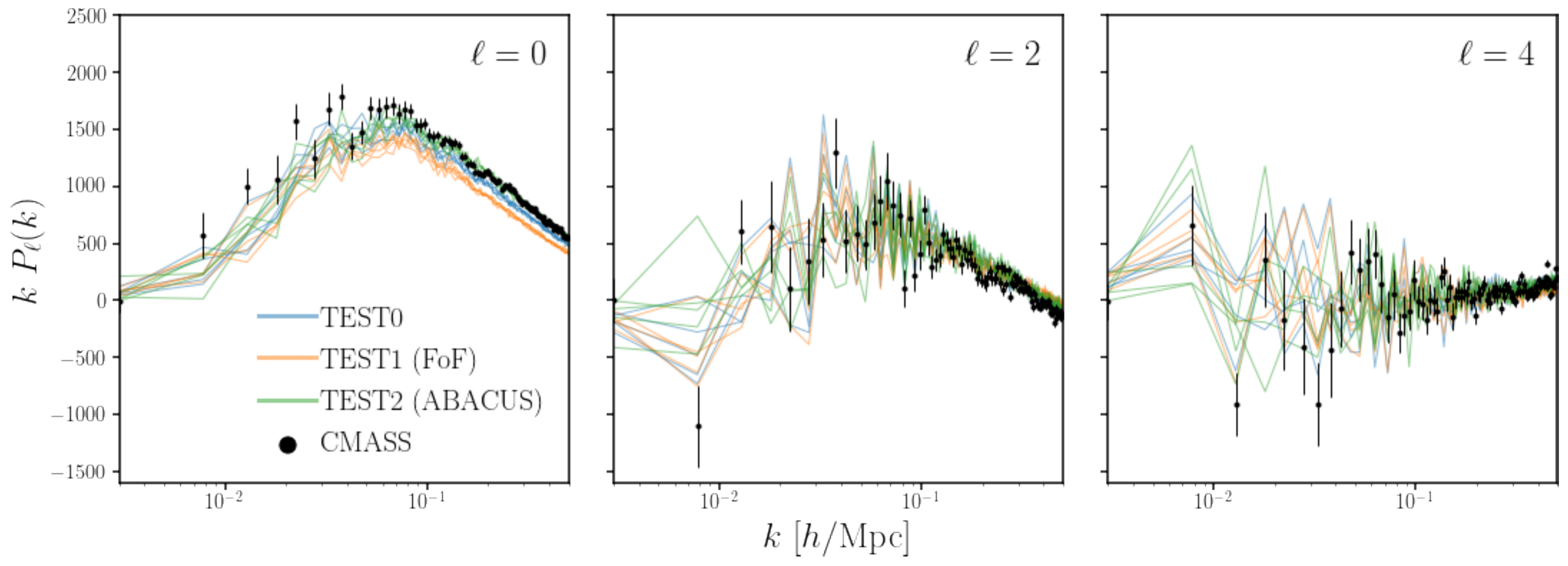}
\caption{
    $k\,P_\ell(k)$ of the three sets of test simulations constructed to validate
    the \simbig~framework.
    In each panel, we present $P_\ell(k)$ of 10 randomly selected simulations 
    from $\mathtt{TEST0}$ (blue), $\mathtt{TEST1}$ (orange), and 
    $\mathtt{TEST2}$ (green). 
    We present $\ell = 0, 2, 4$ in the left, center, and right panels. 
    We include the $k\,P_\ell$ of CMASS galaxies (black) with uncertainties 
    estimated using $\mathtt{TEST0}$ for reference. 
    We construct the test simulations with different sets of assumptions
    ($N$-body simulation, halo finder, HOD model) than the training set and
    thus there are noticeable differences among them. 
    Overall, their $P_\ell$ that loosely agree with the CMASS $P_\ell$. 
    \label{fig:test}
    }\end{center}
\end{figure}

\subsection{{\sc Quijote} Test Simulations} \label{sec:quij_test} 
Our goal for the test simulations is to construct galaxy samples that reflect
the observed CMASS galaxies and are not part of our training data. 
For the first set of test simulations, we use the same forward model as our
training data. 
However, instead of using the $N$-body simulations from the {\sc Quijote}
LHC set, we use a different set of 100 independent {\sc Quijote} simulations 
run at the fiducial cosmology
($\Omega_m = 0.3175, \Omega_b = 0.049, h = 0.6711, n_s = 0.9624, \sigma_8 =
0.834$).
The fiducial simulations have the same properties (\emph{e.g.} volume, resolution)
as the LHC simulations other than cosmology. 
We also sample HOD parameters from a narrower distribution of HOD parameters
than the prior (Section~\ref{sec:prior}).
For the Z07 HOD parameters, we center the range around the best-fit HOD
parameter values from \cite{reid2014} with widths 0.058, 0.12, 0.26, 0.12,
0.36 for $\log M_{\rm min}, \sigma_{\log M}, \log M_0, \log M_1, \alpha$,
respectively. 
We sample 
$A_{\rm bias} \sim \mathcal{N}(0., 0.02)$, 
$\eta_{\rm conc} \sim \mathcal{U}(0.9, 1.1)$, 
$\eta_{\rm cen} \sim \mathcal{U}(0., 0.1)$, and 
$\eta_{\rm sat} \sim \mathcal{U}(0.9, 1.1)$. 
The rest of our forward model (volume remapping, survey geometry, systematics)
is applied in the same way as our training data. 
We use 5 different HOD parameter values per $N$-body simulation for a total of
500 test galaxy catalogs. 
We refer these test simulations as $\mathtt{TEST0}$. 

For the second set, we again construct our test galaxy samples from the
{\sc Quijote} fiducial $N$-body simulations but using a different halo finder
and HOD model. 
Instead of the {\sc Rockstar} halo finder we use halo catalogs constructed
using the Friend-of-Friend (FoF) algorithm~\citep{davis1985} with the linking 
length parameter set to $b=0.2$. 
Furthermore, instead of the HOD model with assembly, concentration, and
velocity biases we use the the Z07 HOD model with no assembly, concentration,
or satellite velocity bias.
We include central velocity bias because the halo velocities in FoF halo
catalogs correspond to the bulk velocity of the dark matter particles in the
halo rather than the velocity of the central density peak of the halo, which
better corresponds to the central galaxy velocity~\citep{knebe2011,
behroozi2013a}.
If we ignore this discrepancy, we produce an imprint of Fingers-of-God (FoG) on
the power spectrum quadrupole that significantly deviates from observations. 
We sample 5 HOD parameters for each $N$-body simulation using the same
range as $\mathtt{TEST0}$ for the Z07 HOD parameters and use fixed
$\eta_{\rm cen} = 0.2$. 
We refer the second set of 500 test simulations as $\mathtt{TEST1}$. 

\subsection{{\sc Abacus} Test Simulations} \label{sec:abac_test} 
In addition to the {\sc Quijote} based test simulations above, we construct a 
third, and most stringent, set of test simulations designed to test the
assumptions in our $N$-body simulations and halo finder. 
We build the test simulations using the {\sc AbacusSummit} $N$-body
simulation~\citep{maksimova2021} and the {\sc CompaSO} halo
finder~\citep{hadzhiyska2022}.
{\sc AbacusSummit} is a suite of large, high-accuracy $N$-body simulations
constructed using the {\sc Abacus} code~\citep{garrison2018, garrison2021}. 
For the test simulations, we use 25 simulations in the ``base'' configuration
out of the 150 simulations at 97 different cosmologies in the suite.
The {\sc AbacusSummit} simulations contain $6912^3$ particles in 
a $(2\,h^{-1}{\rm Gpc})^3$ volume box and have significantly higher resolution 
than {\sc Quijote}. 
% method grove2022 n-body simulation comparison 

From {\sc AbacusSummit}, we use halo catalogs generated using {\sc CompaSO}, 
a specialized spherical overdensity based halo finder. 
{\sc CompaSO} improves halo deblending from spherical overdensity algorithms 
by considering the tidal radius around smaller halos before halo assignment.
It also improves known limitation in identifying halos close to the center of
mass of larger halos. 
{\sc CompaSO} also utilizes a post-processing~\citep{bose2022} step to remove
over-deblended halos and merge physically associated halos that have merged and
then physically separated.
This halo finder has been used with HOD models to accurately fit observed
galaxy clustering~\citep[\emph{e.g.}][]{yuan2022}. 

We use halo catalogs from the simulations of the {\sc AbacusSummit} base
configuration. 
We divide each halo catalog into 8 catalogs in $(1\,h^{-1}{\rm Gpc})^3$ volume 
boxes. 
Afterwards, we construct galaxy catalogs with the same HOD model as our training
data with 5 sets of HOD parameter values per simulation sampled from the same
parameter ranges as $\mathtt{TEST0}$. 
We apply the rest of our forward model in the same way as our training data to
construct realistic CMASS-like test galaxy catalogs. 
We refer the third set of 1000 test simulations as $\mathtt{TEST2}$. 

In Figure~\ref{fig:test}, we present $k\,P_\ell(k)$ of the $\mathtt{TEST0}$ 
(blue), $\mathtt{TEST1}$ (orange), and $\mathtt{TEST2}$ (green) simulations 
that we use to test \simbig. 
We present 10 randomly selected test simulations from each suite and plot 
$\ell = 0, 2, 4$ in the left, center, and right panels. 
For reference, we present $k\,P_\ell(k)$ of CMASS galaxies in black with 
uncertainties estimated from the $\mathtt{TEST0}$ simulations. 
Overall, the $k\,P_\ell$ among the test suite loosely agree with the observed $P_\ell$. 
There are notable differences among them since we construct each of them 
using different forward models.

%% file: results.tex
\section{Results} \label{sec:results}

\begin{figure}
\begin{center}
    \includegraphics[width=\textwidth]{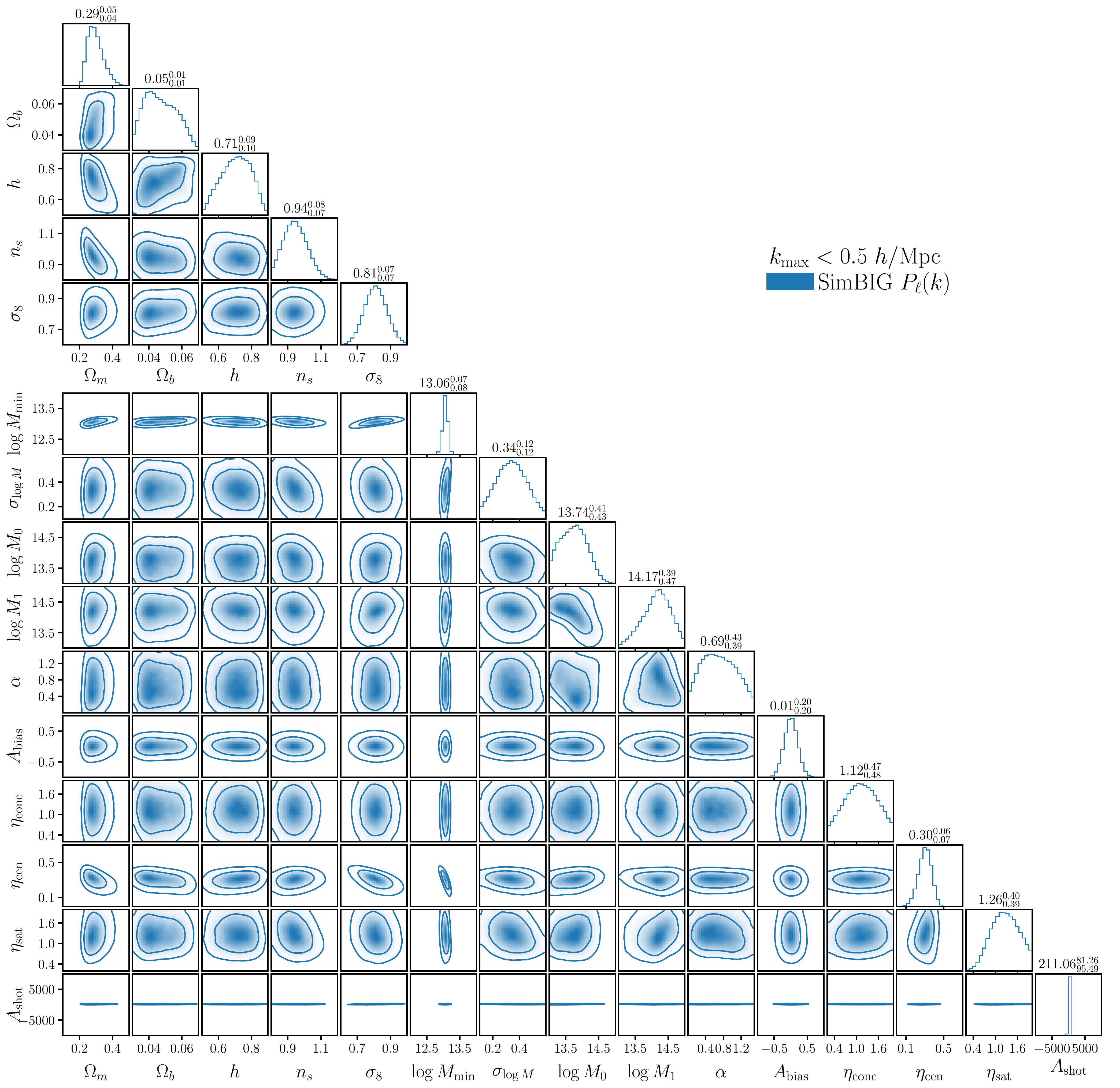}
    \caption{\label{fig:post_full}
    Posterior distribution of all parameters inferred from the BOSS CMASS 
    $P_\ell(k)$ analysis to $k_{\rm max} < 0.5\,h/{\rm Mpc}$ using \simbig.
    In the top set of panels, we present the cosmological parameters. 
    In the bottom, we present the halo occupation and nuisance parameters.
    The range of the panels repersent the prior range in Table~\ref{tab:prior}. 
    Among the halo occupation parameters, the posterior significantly constrains
    $\log M_{\rm min}$ and $\eta_{\rm cen}$. 
    Among the cosmological parameters, the posterior significantly constrains  
    $\Omega_m$ and $\sigma_8$. 
    }
\end{center}
\end{figure}

In Figure~\ref{fig:post_full}, we present the posterior distribution inferred
from the CMASS $P_\ell$ with $k_{\rm max} < 0.5\,h/{\rm Mpc}$ using \simbig. 
In the top panels, we present the posterior of the cosmological parameters.
In the bottom panels, we present the posterior of the halo occupation and
nuisance parameters. 
The diagonal panels present the one-dimensional marginalized posteriors; the
other panels present marginalized posteriors of different parameter pairs. 
The contours represent the 68 and 95 percentiles and the range of the panel
represents the prior range.
We also list the 50, 16, and 84th percentile constraints on the
parameters along the diagonal panels. 

The \simbig~posterior is inferred using mostly uniform priors (except $A_{\rm
bias}$). 
It, therefore, provides insights into which parameters are constrained by $P_\ell$. 
We present and discuss the cosmological constraints in detail in H22a.
Here, we focus on the halo occupation parameters. 
We derive particularly tight constraints on $\log M_{\rm min}$ and 
$\eta_{\rm cen}$. 
Based on the constraints, we find that CMASS central galaxies reside in halos
with $M_h > 10^{13} M_\odot$. 
This is in good agreement with previous CMASS HOD constraints~\citep{reid2014}. 
Our constraints on the Z07 HOD parameters are also in good agreement with 
\cite{kobayashi2021} HOD parameter constraints.
\cite{kobayashi2021} recently analyzed $P_\ell(k)$ using a halo power spectrum
emulator with an analytic prescription for the Z07 HOD model.

Our posterior also provides insights into halo occupation beyond the Z07 model.
For instance, we find a significantly non-zero central velocity bias. 
This suggests that there is a significant velocity offset between CMASS
central galaxies and their halos, which is consistent with past
studies~\citep{guo2015, yuan2020, lange2021, zhai2022}.
Meanwhile, we find little evidence for satellite velocity bias similar
to~\cite{lange2021} and \cite{zhai2022}. 
We also do not find evidence of assembly bias: our posterior does not
significantly constrain $A_{\rm bias}$.
%Among the $\Lambda$CDM cosmological parameters, the \simbig~posterior significantly constrains $\Omega_m$ and $\sigma_8$. 
%This is consistent with previous works~\citep[\emph{e.g.}][]{ivanov2022, kobayashi2022}, where priors from Big Bang nucleosynthesis (BBN) or cosmic
%microwavebackground (CMB) experiments are used for $\Omega_b$ and $n_s$. 

\begin{figure}
\begin{center}
    \includegraphics[width=\textwidth]{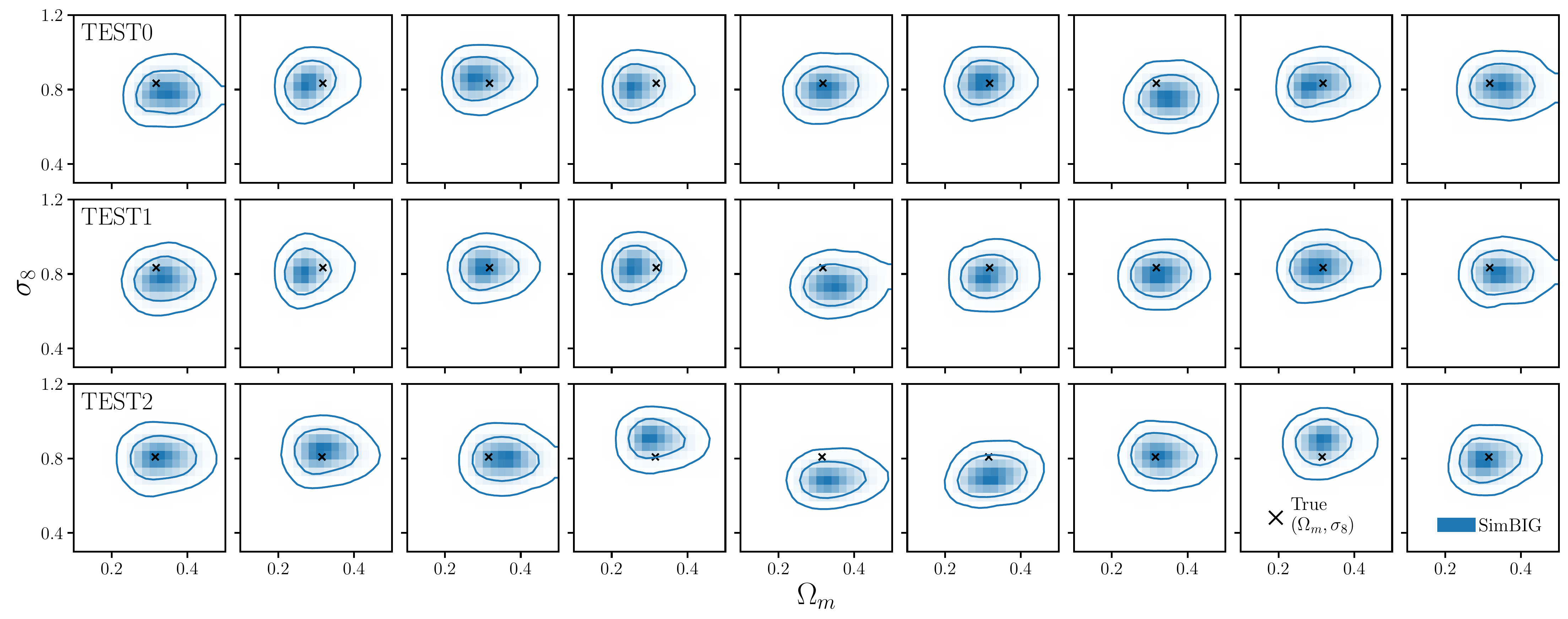}
    \caption{\label{fig:test_post}
    Posteriors of $(\Omega_m, \sigma_8)$ inferred using \simbig~for a random
    subset of the $\mathtt{TEST0}$ (top), $\mathtt{TEST1}$ (center), and 
    $\mathtt{TEST2}$ (bottom) simulations.
    We mark the 68 and 84 percentiles of the posteriors with the contours. 
    We also include the true $(\Omega_m, \sigma_8)$ of the test simulations in
    each panel (black $\times$). 
    The comparison between the posteriors and the true parameter values
    qualitatively show good statistical agreement for each of the test
    simulations. 
    }
\end{center}
\end{figure}

Next, we validate the accuracy of the posterior.  
We run the \simbig~posterior estimator, $q_{\bphi}$, on each of the test simulations
described in Section~\ref{sec:tests}. 
In Figure~\ref{fig:test_post}, we present the posteriors of 
$(\Omega_m, \sigma_8)$ for a randomly selected subset of the test simulations. 
We present posteriors for $\mathtt{TEST0}$, $\mathtt{TEST1}$, and
$\mathtt{TEST2}$ simulations in the top, center, and bottom panels. 
The contours represent the 68 and 95 percentiles of the posteriors. 
In each panel, we mark the true $(\Omega_m, \sigma_8)$ value of the test
simulation (black x). 
Each test simulation is a unique realization of a CMASS-like galaxy catalog
subject to cosmic and statistical variance. 
We, therefore, do not expect the true $(\Omega_m, \sigma_8)$ value to lie at
the center of each of the posteriors. 
Instead, we note that for the majority of the randomly selected test
simulations, the true parameter values lie within the 68 and 95 percentiles
\simbig~posteriors. 

\begin{figure}
\begin{center}
    \includegraphics[width=\textwidth]{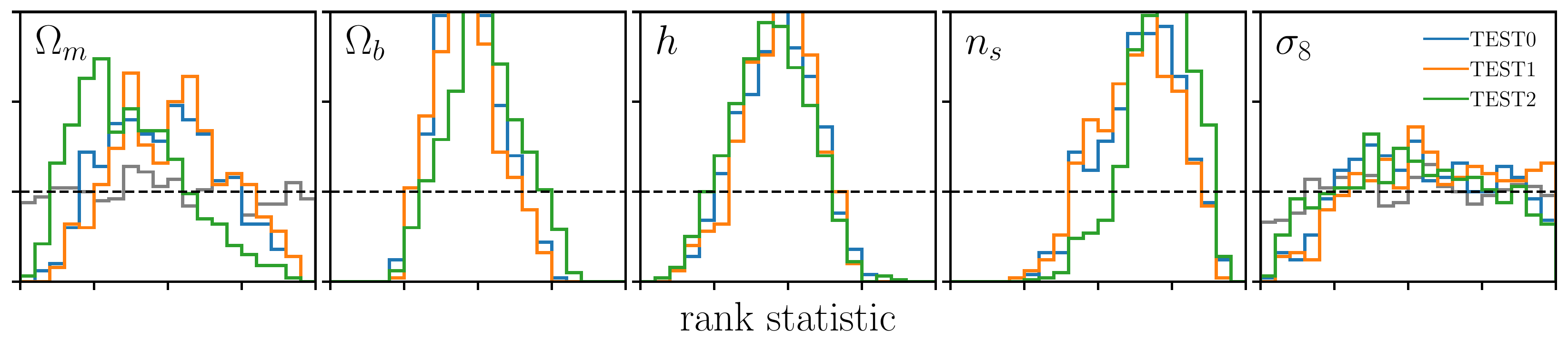}
    \caption{\label{fig:sbc}
    SBC validation of the posteriors estimated using \simbig~for test simulations.
    We present the distribution of the rank statistics, which are derived by
    comparing the true parameter values to the inferred marginalized 1D
    posteriors. 
    In general, for an accurate estimate of the true posterior, the rank statistic
    would be uniformly distributed (black dashed). 
    In our case, since we use test simulations evaluated at fiducial cosmologies, 
    our priors impose a $\cap$-shape, especially for $\Omega_b$, $h$, and $n_s$. 
    We include the expected rank distribution for an accurate estimate of the true 
    posterior for $\Omega_m$ and $\sigma_8$ (gray). 
    Since the distributions for $\Omega_m$ and $\sigma_8$ have symmetric 
    $\cap$-shape with respect to this expected distribution, \simbig~posteriors are 
    unbiased but broader than the true posteriors.
    The rank distributions for different test simulations show an overall good
    agreement and, thus, illustrate that the \simbig~approach is robust to the choices 
    in our forward model.  
    Overall, {\em the \simbig~produces unbiased and conservative posteriors for 
    $\Omega_m$ and $\sigma_8$}.
    }
\end{center}
\end{figure}

We assess the accuracy and precision of the \simbig~more quantitatively
using simulation-based calibration~\citep[SBC;][]{talts2020}. 
For each test simulation and each cosmological parameter, we calculate the rank
of the true parameter value within the marginalized 1D \simbig~posterior.
In practice, for each parameter, $\theta$, and test simulation $j$, we calculate the rank of 
$\theta_{j,{\rm test}}$ within the $\theta$ values sampled from the posterior estimate: 
$\{\theta_1', \theta_2',...,\theta_N'\} \sim q_{\bphi}(\theta|\bfi{x}_{j, {\rm test}}).$
For instance, $\theta_{j, {\rm test}}$ would have rank 1 if 
$\theta_{j, {\rm test}} < \theta_1', \theta_2',...,\theta_N'$
or rank $N$ if $\theta_{j, {\rm test}} > \theta_1', \theta_2',...,\theta_N'$. 
We then compile the ranks for all of the test simulations in each suite and
plot their histogram. 
%examine the distribution of the ranks. 
This distribution and variations of it (conditional coverage) are common 
diagnostics for posterior
estimates~\citep[\emph{e.g.}][]{green2020, hahn2022a, lemos2022}.  
If we estimate the true posterior exactly, the rank statistics would be
distributed uniformly. 
If instead the rank statistics have a U-shaped distribution, the true parameter
values are more often at the lowest and highest ranks, so the posterior
estimates are narrower than the true posteriors.
If the distribution has a $\cap$-shape, then the posterior estimates are
broader than the true posteriors. 
An asymmetric distribution implies that the posterior estimates are
biased. 

In Figure~\ref{fig:sbc}, we present the distribution of rank statistics for the
cosmological parameters in all of the test simulations. 
In each panel, we present the distributions for the $\mathtt{TEST0}$ (blue),
$\mathtt{TEST1}$ (orange), and $\mathtt{TEST2}$ (green) simulations. 
Overall, the distributions have a $\cap$-shape and are symmetric.
The $\cap$-shape is in part due to the prior range on the cosmological parameters 
and the fact that our test simulations are constructed at a fiducial cosmology. 
This is especially the case for $\Omega_b$, $h$, and $n_s$ whose prior ranges
truncate the \simbig~posterior constraints~(Figure~\ref{fig:post_full}). 
The true parameter values of the test simulations are at fiducial values near
the center of the prior: 
$\Omega_b^{\rm fid}, h^{\rm fid}, n_s^{\rm fid} = 0.0490, 0.6711, 0.9624$ for 
$\mathtt{TEST0}$  and $\mathtt{TEST1}$ and 0.0493, 0.6736, 0.9649 for
$\mathtt{TEST2}$.
Hence, since the tails of the likelihood are truncated by the prior, the true 
parameter values will have more central ranks within the marginalized 
posterior. 
Galaxy clustering, however, does not place strong constraints on these 
parameters.
Previous works~\citep{ivanov2020, kobayashi2021} typically use priors from
either big bang nucleosynthesis or CMB constraints for $\Omega_b$ and $n_s$. 
We, therefore, focus on $\Omega_m$ and $\sigma_8$, the cosmological
parameter most significantly constrained by galaxy clustering alone. 

The prior range also affects the rank statistic distribution of $\sigma_8$. 
The $\sigma_8$ posterior is within the prior but has a comparable width to the
prior range (Figure~\ref{fig:post_full}).
To estimate the effect of the prior range on our SBC for $\sigma_8$, we 
consider a simplified scenario assuming a fixed likelihood.
We draw 10,000 samples from the marginalized $\sigma_8$ posterior for an arbitrary 
$\mathtt{TEST0}$ simulation: $\sigma_8' \sim q_{\bphi}(\sigma_8\given\bfi{x}_{\rm test})$. 
The $\mathtt{TEST0}$ simulations have a $\sigma_8$ value of $\sigma_8^{\rm fid}$.
For each sample, $\sigma_8'$, we shift the posterior by 
$\sigma_8^{\rm fid} - \sigma_8'$. 
Since this shift may cause the posterior to go beyond the $\sigma_8$ prior 
of our analysis, we impose the $\sigma_8$ prior range on the shifted posterior. 
Afterward, we calculate the rank of the sample within the truncated and 
shifted posterior.  
We include the rank distribution we obtain from this procedure (gray) for 
both $\Omega_m$ and $\sigma_8$ in Figure~\ref{fig:sbc}.
The suppression of the $\sigma_8$ rank distributions at low ranks is due to
the prior range. 
Once we take the prior range into account, the $\sigma_8$ rank distributions 
for $\mathtt{TEST0}$, $\mathtt{TEST1}$, and $\mathtt{TEST2}$ are in good
agreement with the expected rank distribution for an accurate estimate
of the true posterior. 
We therefore conclude that the \simbig~posterior of $\sigma_8$ is unbiased and
slightly conservative. 

For $\Omega_m$, the posterior is well within the prior so the prior range
does not have a significant impact on its rank statistic distribution 
(see gray distribution in left most panel of Figure~\ref{fig:sbc}). 
Since the prior range does not fully account for the $\cap$-shape, this 
implies that the \simbig~posteriors of $\Omega_m$ is significantly broader 
than the true posteriors. 
This is due to the fact that we use a limited number of training simulations. 
Although we construct 20,000 training simulations, they only sample 2,000
different cosmologies. 
Consequently, our estimate of the KL divergence between the normalizing flow
and the true posterior, which we minimize to train the flow, is intrinsically 
noisy.
So the divergence cannot be further minimized to better estimate the posterior. 
Despite being conservative, the rank statistic distributions are symmetric so 
the \simbig~posterior of $\Omega_m$ is unbiased. 

Figure~\ref{fig:sbc} also reveals the consistency among the rank statistics
distributions for the different test simulations.
We use different forward models to construct $\mathtt{TEST1}$ and
$\mathtt{TEST2}$ than the training data.  
The accuracy and precision of the \simbig~posteriors are not impacted by these
differences and, thus, demonstrates the robustness of our \simbig~approach.
We therefore conclude that {\em we can use \simbig~to infer unbiased and
conservative posteriors from $P_\ell(k)$ to $k_{\rm max} = 0.5\,h/{\rm Mpc}$}. 

%% file: discuss.tex
\section{Discussion} \label{sec:discuss}
Now that we have validated the robustness of the \simbig~posteriors using the
test simulations, we discuss other caveats of the \simbig~framework as well as
improvements for future applications. 
In particular, we focus on the forward model. 
SBI, and thus \simbig, relies on a forward model that can accurately model 
the observables. 

%%%%%%%%%%%%%%%%%%%%%%%%%%%%%%%%%%%%%%%%%%%%%%%%%%%%%%%%%%%%%%%%%%%%%%%%%%%
% fowrad model 
%%%%%%%%%%%%%%%%%%%%%%%%%%%%%%%%%%%%%%%%%%%%%%%%%%%%%%%%%%%%%%%%%%%%%%%%%%%
\subsection{Forward Model} \label{sec:forward-model}
Our forward model consists of the {\sc Quijote} $N$-body simulations, 
the {\sc Rockstar} halo finder, an HOD model, and the model for  BOSS CMASS-SGC
survey realism.  
In this section we examine each aspects of our forward model. 

\subsubsection{$N$-body Simulation} 
We use high resolution {\sc Quijote} $N$-body simulations to model the
clustering of matter. 
Full $N$-body simulations more accurately model non-linear matter clustering
than more approximate methods such as particle mesh
scheme~\citep[\emph{e.g.}][]{feng2016}.
While the accuracy of $N$-body simulations depends on their resolution, the
matter clustering of {\sc Quijote} simulations are converged at
$k\sim0.5\,h/{\rm Mpc}$ even for simulations with lower resolution than our
$1024^3$ resolution~\citep{villaescusa-navarro2020}.
The {\sc Quijote} simulations are run using the TreePM {\sc Gadget}-III code. 
We do not expect the choice of $N$-body code to impact our results as different
$N$-body codes produce highly consistent matter clustering~\citep{shao2022}.  

We also note that the {\sc Quijote} $N$-body simulations do not include
baryonic effects. 
Feedback from active galactic nuclei (AGN), for instance, can impact the matter
distribution at cosmological distance~\citep{vandaalen2011, vogelsberger2014,
hellwing2016, peters2018, springel2018, chisari2018, barreira2019, foreman2019,
vandaalen2020}. 
The impact, however, is mainly found on very small scales and is a subpercent
effect on the power spectrum at $k < 0.5\,h/{\rm Mpc}$. 
We note that other baryonic processes also take effect on smaller scales than
our analysis~\citep[\emph{e.g.}][]{white2004, zhan2004, jing2006, rudd2008,
harnois-deraps2015}. 
We, therefore, do not expect baryonic effects to have a significant impact on
our parameter constraints. 

\subsubsection{Halo Occupation} 
From the matter distribution, we simulate the galaxy distribution using the
{\sc Rockstar} halo finder and an HOD model. 
{\sc Rockstar} is a phase-space based halo finder designed to maximize halo
consistency across time-steps. 
It first selects particle groups using a 3D FoF algorithm with a large linking
length, then builds a hierarchy of FoF subgroups in phase space by
progressively and adaptively reducing the linking length. 
Afterwards, {\sc Rockstar} converts the FoF subgroups into halos starting from
the deepest level of the hierarchy. 
According to the \cite{knebe2011} comparison of 18 different halo finders
using test halo simulations, phase-space halo finders like {\sc Rockstar} can
accurately resolve the spatial location of halos as well as the substructure
near the center of halos. 
Furthermore, {\sc Rockstar} can accurately resolve substructure containing 10-20 particles. 
\cite{knebe2011} found that phase-space halo finders are in good agreement for
halo properties. 
They, however, found significant discrepancies among the halo finders on subhalo
properties.
In our forward model, we only use central halos so we are not impacted by the
lack of convergence in subhalo properties among halo finders. 
%Since we only use central halos, we are also not impacted by other issues such as unresolved ``orphan'' subhalos (\todo{cite}). 

The HOD model that we use to populate halos with galaxies includes assembly
bias, concentration, central velocity, and satellite velocity biases. 
This is a state-of-the-art HOD model that provides a highly flexible framework
for populating galaxies in halos. 
As we demonstrate above, it is sufficiently flexible to reproduce the
observations for $P_\ell(k)$. 
However, this may not be the case for summary statistics beyond the power 
spectrum that may be more sensitive to the limitations of the HOD model. 
Whether the HOD model is sufficiently descriptive must be determined for each
summary statistics. 

\subsubsection{Survey Realism} 
The last step of our forward model is to apply the survey realism of the BOSS
CMASS SGC sample.
From the galaxy distribution in a $(1\,h^{-1}{\rm Gpc})^3$ box, we cut out the
exact survey volume of the CMASS sample. 
One detail omitted in this procedure is the redshift evolution of the galaxy
distribution. 
Our galaxy distribution is constructed at a single $z=0.5$ snapshot. 
However, we do not expect a significant redshift dependence in the galaxy
distribution or the underlying matter distribution over our narrow redshift
range, $0.45 < z < 0.6$. 
We also do not expect a significant redshift dependence on the HOD, since CMASS
galaxies were selected to have a roughly constant mass limit throughout the 
redshift range. 

After we apply the survey geometry, we impose observational systematics. 
A key systematic that we include in our forward model is the effect of fiber
collisions. 
For our forward model, we ``collide'' galaxies by randomly select 60\% of
galaxy pairs that have angular separation less than 62\arcsec and removing one
of the galaxies in the selected pair.
In this implementation, fiber collisions occur uniformly over the survey
footprint. 
In principle, fiber collisions in SDSS-III depend on the tiling scheme used and
are not uniformly distributed~\citep{dawson2013}.
Regions observed by overlapping tiles have lower fiber collision rates than
regions observed by a single tile~\citep{guo2012}. 
\cite{hahn2017a} examined the impact of fiber collisions on $P_\ell(k)$ using
two sets of simulations Nseries and QPM~\citep{white2014}. 
Nseries applied fiber collisions using the full SDSS tiling scheme while QPM
applied fiber collisions in the same fashion as our forward model.
\cite{hahn2017a} examined the impact of fiber collisions on $P_\ell$ for both
of the simulations and found little difference.
We, therefore, conclude that the angular dependence of fiber collisions is not
a significant effect for our analysis.  

Another observational effect that we do not include in our forward model is
incompleteness. 
The CMASS sample does not include all galaxies within its angular footprint
that pass the target selection criteria.  
There is incompleteness due to imaging systematics. 
CMASS galaxies are selected for observation using the SDSS imaging data. 
Systematics in the imaging, such as seeing, sky background, airmass,
galactic extinction, have significant correlations with the number density of
galaxies.
There is also incompleteness due to failures in measuring an accurate redshift. 
Redshift failures occur more frequently for galaxy spectra measured using a
fiber near the edge of the focal plane. 
\cite{ross2012} and \cite{ross2017} model these effects and derive weights
that can accurately correct for them.
We opt to use these completeness weights since they have been demonstrated 
to sufficiently correct for incompleteness in the two-point clustering. 
Yet they may not be sufficient for higher-order
We reserve a more detailed investigation of the impact of completeness to
future work. 

%%%%%%%%%%%%%%%%%%%%%%%%%%%%%%%%%%%%%%%%%%%%%%%%%%%%%%%%%%%%%%%%%%%%%%%%%%%
% outlook 
%%%%%%%%%%%%%%%%%%%%%%%%%%%%%%%%%%%%%%%%%%%%%%%%%%%%%%%%%%%%%%%%%%%%%%%%%%%
\subsection{Outlook} \label{sec:outlook} 
In this work, we present the mock challenge validation framework and additional details of \simbig.
We demonstrate that we can use \simbig~to analyze the BOSS CMASS galaxy sample
with $P_\ell$ as our summary statistics. 
This is only the first steps. 
The \simbig~framework can be applied beyond the BOSS CMASS $P_\ell$. 

\subsubsection{Beyond $P_\ell(k)$}
In subsequent work, we will use \simbig~to analyze summary statistics that
forecasts show can significantly improve cosmological constraints over the
power spectrum~\citep{hahn2021a, massara2022,hou2022}.  
We will analyze the BOSS galaxy bispectrum, the simplest statistic beyond
$P_\ell$ that captures non-Gaussian galaxy clustering information
(Hahn~\etal~in prep.).
We will also analyze the marked power spectrum multipoles, which measures the
two-point clustering statistics of a weighted galaxy field (Massara~\etal~in
prep.).
In another work, we will present constraints from the weighted skew spectra,
which are simple and interpretable proxy statistics for the bispectrum
(Hou, Moradinezhad~\etal~in prep.).

The \simbig~framework also enables clustering analyses using more novel summary
statistics. 
For instance, \cite{eickenberg2022} recently demonstrated that 3D wavelet
statistics can extract non-Gaussian information and significantly improve
cosmological parameter constraints.   
We will present cosmological constraints from BOSS CMASS using using wavelet
statistics in Eickenberg, Regaldo,~\etal~(in prep.).
With \simbig~we can analyze any summary statistic that can be measured in the 
observed galaxy distribution. 
Hence, we will also present constraints from summary statistics that compress 
the full BOSS CMASS galaxy field using convolutional and graph NNs~(Lemos~\etal~in prep.).

\subsubsection{Beyond BOSS CMASS}
While all of the studies described above will analyze the BOSS CMASS sample,
\simbig~can be extended to upcoming and future surveys such as DESI, PFS,
{\em Euclid}, and Roman. 
In particular, the DESI Bright Galaxy Survey~\citep[BGS;][]{hahn2022c} and PFS
will provide high density samples ideal for \simbig-like analyses.
BGS will observe a $r < 19.5$ magnitude-limited galaxy sample out to $z=0.6$
over a 14,000 ${\rm deg}^2$ angular footprint. 
It will also have a number density an order-of-magnitude higher than CMASS.
Meanwhile, PFS will observe a high number density sample of emission line
galaxies (ELGs) over the range $0.6 < z < 2.4$ over a 1,200 ${\rm deg}^2$
angular footprint. 
In this work, we analyze the BOSS CMASS-SGC sample because of the volume
and resolution limit of the {\sc Quijote} $N$-body simulations. 
Both BGS and PFS will cover substantially larger volumes.
This will be a challenge in extending \simbig. 

Fully spanning the BGS and PFS volume will require significantly larger
simulations of 3 and $7\,(h^{-1}{\rm Gpc})^3$, respectively. 
Both BGS and PFS galaxy samples will also require significantly higher
resolution than {\sc Quijote}. 
BGS will probe galaxies with stellar masses as low as $M_* \sim 10^{7}M_\odot$.
Meanwhile, ELGs can reside in low mass halos with $M_h \sim 10^{11} M_\odot$. 
Fortunately, new developments in computational techniques and ML will make it
possible to construct larger and higher-resolution simulations. 
Approximate $N$-body schemes, such as particle mesh, are
significantly improving in both accuracy and speed~\citep{feng2016, modi2021}. 
Their accuracy can also be enhanced using methods such as potential gradient
descent~\citep{dai2020a}.

Furthermore, ML methods have now firmly demonstrated that they can be used to
efficiently construct accurate high-resolution simulations. 
\cite{schaurecker2021}, for example, showed that convolutional NN can be used
to enhance low resolution $N$-body simulations to create super-resolution
simulations that accurately reproduce the cluster of high resolution
simulations. 
Along similar lines, \cite{alvesdeoliveira2020} and \cite{jamieson2022}
successfully constructed accurate emulators for high resolution simulations. 
These ML methods would enable us to leverage a limited number of simulations, 
used for training, to construct a large set of emulated simulations with 
comparable accuracy. 

\subsubsection{Beyond $\Lambda$CDM}
We assume a $\Lambda$CDM cosmology for the cosmological constraints that we
present in this work.  
The subsequent \simbig~papers will also focus on $\Lambda$CDM cosmological 
constraints.
With \simbig, however, we can incorporate cosmological models beyond
$\Lambda$CDM. 
We can expand the cosmological parameters to include, for instance, the dark
energy equation of state, massive neutrinos, or primordial non-Gaussianity
by replacing the $N$-body simulations. 
All other aspect of the \simbig~forward modeling pipeline does not depend on
cosmology. 

We expect the \simbig~approach to be especially effective for measuring the 
total mass of neutrinos, $M_\nu$. 
Recent forecasts illustrate that higher-order statistics can tightly constrain
$M_\nu$~\citep[\emph{e.g.}][]{hahn2020, massara2020, hahn2021a}. 
Furthermore, massive neutrinos suppress the growth of structure on small scales
below their free-streaming scale so $M_\nu$ analyses require careful treatment 
of any systematics that affect small scale clustering. 
\simbig~provides a comprehensive framework for exploiting higher-order
galaxy clustering down to small scales.  
%In \simbig, we include observational systematics in the forward model.  This provides a more direct way of modeling and accounting for systematics.

\subsubsection{Beyond the HOD}
A major ingredient of our forward model is the HOD, which provides the mapping
between the dark matter halos and galaxy distributions. 
Halo occupation in current HOD models only depends on a few halo properties
(\emph{e.g.} halo mass). 
There is, however, growing evidence that the galaxy-halo connection depends
significantly on the halos' detailed assembly
history~\citep[\emph{e.g.}][]{gao2005, wechsler2006, gao2007, zentner2007,
dalal2008, lacerna2011, miyatake2016, more2016, zentner2016, lehmann2017, 
vakili2019, hadzhiyska2021, hadzhiyska2022a}.
This effect is commonly referred to as ``halo assembly bias'' and we include it
in our HOD model by include a dependence of the galaxy occupation on halo
concentration. 
Although we do not find significant constraints on assembly bias 
($A_{\rm bias}$), this is not evidence of the lack of assembly bias.  
Assembly bias may yet have a significant impact on more precise galaxy
clustering measurements.
It may also impact summary statistics beyond $P_\ell$.  
Moreover, our decorated HOD implementation may not accurately model the
effect of assembly bias in detail. 

Fortunately, there are various avenues for improving halo occupation models.  
For instance, \cite{delgado2022} demonstrated that halo occupation can be more
accurately predicted if information on local environmental overdensity and
shear is included. 
\cite{jespersen2022} similarly find that galaxy properties, and
thus occupation, can be more accurately predicted by including information on
halo assembly history. 
Moreover, halo occupation models encapsulate only a particular aspect of galaxy
formation and evolution. 
Better understanding of galaxies will, therefore, improve halo occupation
models.  
We emphasize that \simbig, with its modular forward modeling approach, provides
an ideal cosmological analysis framework for incorporating such improved
models.

Along these lines, the \simbig~approach also provides steps towards more fully
extracting the cosmological information from galaxies and their detailed
properties. 
\cite{villaescusa-navarro2022} controversially claimed that there may be 
significant cosmological information even in a single galaxy. 
The precision for which we can measure the detailed properties of galaxies such
as stellar mass and star formation rate will pose major challenges for any
cosmological inference with a single galaxy.
However, cosmological inference may be possible with statistically powerful
galaxy samples with measured galaxy properties. 
For instance, the probabilistic value-added BGS catalog~\citep{hahn2022} will
provide stellar mass, star formation rate, and stellar metallicity measurements
of 15 million BGS galaxies. 

We can in principle replace the halo occupation model in \simbig~with galaxy
formation models that predict galaxy properties, such as semi-analytic models
or emulators of cosmological hydrodynamical simulations~\citep[see ][for a
review]{somerville2015a}.
Then, with the SBI framework of \simbig, we can conduct robust cosmological
inference using summary statistics that take advantage of galaxy properties.
Ultimately, \simbig~provides a framework for maximally extracting
cosmological information from spectroscopic galaxy surveys. 

%% file: summary.tex
\section{Summary} \label{sec:summary}
We present \simbig, a forward modeling framework for constraining cosmological
parameters from galaxy clustering using simulation-based inference. 
\simbig~enables robust analysis of galaxy clustering on non-linear scales with 
higher-order statistics, that accounts for observational systematics.  
As a demonstration, we use \simbig~to analyze $P_\ell(k)$ of the SDSS-III BOSS CMASS SGC
galaxy sample to $k_{\rm max} = 0.5\,h/{\rm Mpc})$.
In an accompanying paper~\citep{simbig_letter}, we present the cosmological
constraints in detail. 
In this work, we describe our forward model and present the mock challenge for 
validating our cosmological parameter constraints using a suite of test simulations. 

% forward model paragraph 
Our forward model is designed to model the observed galaxies in the CMASS SGC 
sample.
It is based on 2000 high-resolution {\sc Quijote} $N$-body simulations that
are evaluated at different cosmologies arranged in a latin hypercube
configuration to impose uniform priors. 
From the $N$-body simulations we construct galaxy simulations by first
identifying dark matter halos using {\sc Rockstar} and then populating the halos
using a flexible HOD model. 
Our HOD model extends the standard \cite{zheng2007} model by including 
assembly, concentration, central velocity, and satellite velocity biases. 
As the last step of our forward model, we apply full survey realism onto the
galaxy simulations.
We model the BOSS CMASS-SGC survey geometry and apply angular masking as well
as fiber collisions.
In total we construct 20,000 forward model galaxy catalogs.
We measure $P_\ell(k)$ of these catalogs and use the measurements to train our
normalizing flow and conduct SBI. 

% mock challenge set up summary
For our mock challenge, we construct three sets of test simulations: $\mathtt{TEST0}$,
$\mathtt{TEST1}$, and $\mathtt{TEST2}$. 
$\mathtt{TEST0}$ is constructed using the same forward model as the
simulations used for SBI but with {\sc Quijote} simulations at a fiducial
cosmology.  
$\mathtt{TEST1}$ is constructed using a different forward model. 
We use the same $N$-body simulations but with a different halo finder (FoF) and
HOD model. 
$\mathtt{TEST1}$ is also constructed using a different forward model, where we
use a different $N$-body simulation ({\sc AbacusSummit}) and a different halo
finder ({\sc CompaSO}). 
To validate \simbig, we infer posteriors for $P_\ell(k)$ measured for each of
the test simulations. 
We then use the true cosmologies of the test simulations to assess the accuracy
and precision of the \simbig~posteriors.  

Based on the \simbig~posteriors and the mock challenge, we find the following
results.  
\begin{itemize}
    \item From our $P_\ell$ \simbig~analysis, we derive constraints on cosmological,
    HOD, and nuisance parameters. 
    Among the HOD parameters, our constraints on the Z07 model parameters are in 
    good agreement with previous works~\citep{reid2014, kobayashi2021}. 
    Our constraints on central and satellite velocity biases are also in good agreement
    with the literature~\citep{guo2015, yuan2020, lange2021, zhai2022}.
    We do not significantly constrain assembly bias.
    For our cosmological constraints, we refer readers to H22a.

    \item Based on the mock challenge, we demonstrate that \simbig~posteriors
    are unbiased. 
    \simbig~infers statistically unbiased posteriors of the cosmological
    parameters for the test simulations. 
    Since the test simulations are constructed using significantly different
    forward models, this is a clear demonstrate of the robustness of \simbig.

    \item We also find that the \simbig~posteriors are conservative. 
    We derive broader posteriors than the true posteriors for the test simulations.
    This is due to the limited number of $N$-body simulations used in our SBI.
    Our training dataset is based on only 2,000 different cosmological parameters.
    Additional simulations will improve the accuracy of the \simbig~posteriors. 
\end{itemize}

Overall, the mock challenge results demonstrate that we can use \simbig~to robustly 
and accurately analyze galaxy clustering.
Although our posteriors are conservative, because we extract cosmological
information on small scales inaccessible with perturbation theory analyses, our
constraints are competitive. 

In this work and \cite{simbig_letter}, we analyze $P_\ell(k)$ for the primary
purpose of demonstrating the \simbig~framework. 
Our $P_\ell(k)$ analysis only takes advantage of Gaussian cosmological information
down to small scales.
Forecasts suggest that significant amount of non-Gaussian cosmological information 
can be extracted using higher-order statistics. 
In subsequent work, we will use \simbig~to analyze observables beyond $P_\ell$,
including the bispectrum, marked power spectrum, skew spectra, wavelet
statistics, and field-level statistics.

The \simbig~framework is currently designed to analyze the BOSS CMASS galaxies. 
However, \simbig~provides a highly modular framework that can be used to
analyze upcoming galaxy surveys. 
The \simbig~forward model can be enhanced, for instance, by replacing the {\sc
Quijote} simulations with higher resolution or larger simulations. 
New halo occupation models being developed in the literature can also improve
how we populate the forward model with galaxies. 
New applications of ML also offer many opportunities to improve \simbig~by
exploiting fast emulation or super-resolution. 
\simbig~will be particularly effective for analyzing the upcoming DESI BGS and
PFS that will provide high density galaxy samples over unprecedented cosmic
volumes.